# Electronic structure investigation of the cubic inverse perovskite Sc$_3$AlN

Martin Magnuson[1,*], Maurizio Mattesini[2], Carina Höglund[1], Igor A. Abrikosov[1], Jens Birch[1] and Lars Hultman[1]

[1]*Department of Physics, IFM, Thin Film Physics Division, Linköping University, SE-58183 Linköping, Sweden.*
[2]*Departamento de Física de la Tierra, Astronomía y Astrofísica I, Universidad Complutense de Madrid, E-28040, Spain.*

**Abstract**

The electronic structure and chemical bonding of the recently discovered inverse perovskite Sc$_3$AlN, in comparison to ScN and Sc metal have been investigated by bulk-sensitive soft x-ray emission spectroscopy. The measured Sc *L*, N *K*, Al $L_1$, and Al $L_{2,3}$ emission spectra are compared with calculated spectra using first principle density-functional theory including dipole transition matrix elements. The main Sc 3*d* - N 2*p* and Sc 3*d* - Al 3*p* chemical bond regions are identified at -4 eV and -1.4 eV below the Fermi level, respectively. A strongly modified spectral shape of 3*s* states in the Al $L_{2,3}$ emission from Sc$_3$AlN in comparison to pure Al metal is found, which reflects the Sc 3*d* - Al 3*p* hybridization observed in the Al $L_1$ emission. The differences between the electronic structure of Sc$_3$AlN, ScN, and Sc metal are discussed in relation to the change of the conductivity and elastic properties.

## 1 Introduction

The early transition metal nitrides have great technological potential as hard coatings and wide band gap semiconductors and possibly for applications in magnetic recording and sensing. However, Sc-based ternary nitride compounds are largely unexplored. The family of compounds with perovskite crystal structure is of scientific interest due to their versatile physical properties such as high Youngs modulus, high melting point, colossal magnetoresistance (CMR), and in some cases high-temperature superconductivity (HTS), and ferroelectricity. One of the most stable crystal structures for ternary and multinary compounds is the *cubic* perovskite. Recently, Sc$_3$AlN, a cubic *inverse* perovskite phase in the ternary Sc-Al-N system, has been synthesized in our laboratory [1]. Presently, there exist only a very limited number of perovskites with cubic inverse perovskite crystal structure, for example, Ti$_3$AlN [2] and Sc$_3$InN [3]. Previous experimental investigations of the occupied and unoccupied electronic states of ternary nitride Ti-Al-N systems include nanolaminated Ti$_2$AlN [4]. Recently, a stable Sc$_3$GaN cubic inverse perovskite phase was also theoretically predicted [5].

The cubic inverse perovskite nitrides constitute a relatively unknown branch of the perovskite family. Thus, their electronic structure and chemical bonding have not previously been investigated. The fundamental mechanism for chemical bonding in specific crystal directions leading to certain macroscopic phenomena as high conductivity and elasticity or stiffness needs to be further understood. Detailed knowledge about the character of the electronic states in the interior of inverse perovskites is the key for this understanding. Further, it is important to compare spectra between perovskites and inverse perovskites to understand how the differences in the electronic structure leads to specific physical properties.

The inverse perovskite is a cubic crystal structure with light metal atoms (Al) at the cube corner sites, heavy metal atoms (Sc) at face centered positions, and the non-metal atoms (N) in the body centered position. In effect, the inverse perovskite is an ordinary perovskite where the heavy metal atoms have exchanged positions with the non-metal atoms within the unit cell, as shown in Fig. 1.

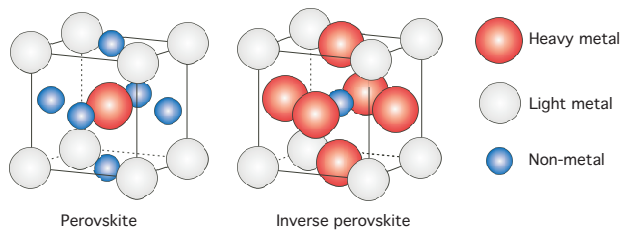

**Figure 1:** (Color online) The crystal structures of perovskites and *inverse* perovskites. The heavy metal (Sc) and non-metal (N, C or O) atoms have exchanged positions with each other in the inverse perovskite (Sc$_3$AlN) in comparison to the perovskite structure. The measured (calculated) cubic lattice parameter *a* is 4.40Å (4.42Å) in Sc$_3$AlN. For comparison, the measured (calculated) lattice parameter for fcc ScN *a* is 4.50Å (*a*=4.54Å) and for hcp Sc *a*=3.31 and *c*=5.27Å (*a*=3.33Å and *c*=5.30Å).

Here, we investigate the internal electronic structure and the influence of hybridization and chemical bonding between the constituent atoms of the cubic inverse perovskite Sc$_3$AlN in comparison to semiconducting fcc ScN and hcp Sc metal, using bulk-sensitive and element-specific soft x-ray emission (SXE) and soft x-ray absorption (SXA) spectroscopy of single-crystal thin film samples. Experimental and calculated SXA and SXE spectra of pure Sc metal are presented for the first time. The SXE technique - with selective excitation energies around the Sc 2*p*, N 1*s*, Al 2*s*, and Al 2*p* absorption thresholds - is more bulk sensitive than electron-based spectroscopic techniques. Due to the involvement of both valence and core levels, the corresponding difference in energies of the emission lines and their dipole selection rules, each kind of atomic element can be probed separately. This enables us extract both elemental and chemical bonding information of the electronic structure of the valence bands. The SXE and SXA spectra are interpreted in terms of partial density of states (pDOS) weighted by the dipole transition matrix elements.

## 2 Experimental

### 2.1 Deposition of Sc$_3$AlN, ScN, and Sc films

Single crystal, stoichiometric Sc$_3$AlN(111), ScN(111), and Sc(0001) thin film samples were grown epitaxially by reactive magnetron sputtering onto polished MgO(111) substrates, 10×10×0.5 mm$^3$ in size, at 650, 600, and 280$^o$C, respectively. The depositions were carried out in an ultrahigh vacuum chamber with a base pressure of 1×10$^{-8}$ Torr from unbalanced type II magnetrons with 50 mm diameter Sc and 75 mm diameter Al targets. Prior to deposition, the substrates were cleaned in ultrasonic baths of trichloroethylene, acetone and isopropanol, and blown dry in dry N$_2$. This was followed by degassing in the chamber for 1 h at 900$^o$C before ramping down to the deposition temperature. The substrate potential was set to be floating. Prior to the deposition of Sc$_3$AlN, the substrate was covered with a ScN(111) seed layer to avoid interdiffusion between the film and the substrate. During deposition, the Ar and N$_2$ partial pressures were kept at 8.0 mTorr/0.2 mTorr, 8.0 mTorr/0.3 mTorr, and 5.0 mTorr/0 mTorr for Sc$_3$AlN, ScN, and Sc, respectively, while the Sc and Al magnetron powers were set to 250 W/70 W, 200 W/0 W, and 100 W/0 W, respectively. The thicknesses of the Sc$_3$AlN, ScN and Sc films were 1750 nm, 200 nm and 200 nm, respectively [1]. The crystal structure was characterized with Θ/2Θ diffraction (XRD) scans in a Philips Bragg-Brentano diffractometer using Cu-K$_\alpha$ radiation shown in Fig. 2.

## 2.2 X-ray emission and absorption measurements

The SXE and SXA measurements were performed at the undulator beamline I511-3 at MAX II (MAX-lab National Laboratory, Lund University, Sweden), comprising a 49-pole undulator and a modified SX-700 plane grating monochromator [6]. The SXE spectra were measured with a high-resolution Rowland-mount grazing-incidence grating spectrometer [7] with a two-dimensional multichannel detector with a resistive anode readout. The Sc $L$ and N $K$ SXE spectra were recorded using a spherical grating with 1200 lines/mm of 5 m radius in the first order of diffraction. The Al $L_1$ and $L_{2,3}$ spectra were recorded using a grating with 300 lines/mm, of 3 m radius in the first order of diffraction. The SXA spectra at the Sc $2p$ and N $1s$ edges were measured at normal incidence with 0.08 eV resolution using total electron yield (TEY) and fluorescence yield (TFY), simultaneously. During the Sc $L$ - N $K$, Al $L_1$, $L_{2,3}$ SXE measurements, the resolutions of the beamline monochromator were 0.45, 0.2 and 0.1 eV, respectively. The Sc $L$ - N $K$, Al $L_1$ and Al $L_{2,3}$ SXE spectra were recorded with spectrometer resolutions of 0.42, 0.3 and 0.06 eV, respectively. All measurements were performed with a base pressure lower than $5\times10^{-9}$ Torr. In order to minimize self-absorption effects [8], the angle of incidence was $20^o$ from the surface plane during the emission measurements. The x-ray photons were detected parallel to the polarization vector of the incoming beam in order to minimize elastic scattering.

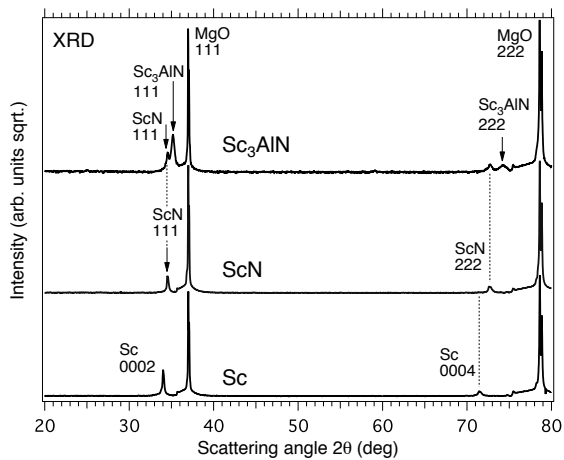

**Figure 2:** X-ray diffractograms (XRD) from the Sc$_3$AlN(111), ScN(111), and Sc(0001) thin film samples.

## 3 *Ab initio* calculation of soft x-ray emission spectra

The SXE spectra were calculated with the WIEN2k code [9] employing the density-functional [10, 11] augmented plane wave plus local orbital (APW+lo) computational scheme. The APW+lo method expands the Kohn-Sham orbitals in atomic-like orbitals inside the muffin-tin (MT) atomic spheres and plane waves in the interstitial region. For each particular atomic arrangement (*hcp*-Sc, *fcc*-ScN and *sc*-Sc$_3$AlN), the Kohn-Sham equations were solved using the Wu-Cohen generalized gradient approximation (GGA) [12, 13] for the exchange-correlation potential.

In all the analyzed structures we used a muffin-tin radius (R$_{MT}$) of 2.20 atomic units (a.u.) for both Sc and Al, while for N a value of 1.90 a.u. was fixed. The Sc $1s^2$, $2s^2$ and $2p^6$ states were considered as core states, and they were treated using only the spherical part of the potential. In the same manner we handled the Al $1s^2$ and $2s^2$ states and the N $1s^2$ electrons. Only when calculating the Al $L_{2,3}$-edges we set up the aluminium $2p^6$ semi-core electrons as pure core states and used an energy cut-off of -4.5 Ry for separating core from valence states, thus leaving the $3s^2$ and $3p^1$ configuration into the valence shell. For the calculation of the valence part, we considered an expansion of the potential and the charge density into spherical harmonics up to $\ell=4$. The valence wave functions inside the atomic spheres were expanded up to $\ell=10$ partial waves. For Sc $s$, $p$ and $d$ local orbitals were added to the APW basis set to improve the convergence of the wave function while, for N and Al only $s$ and $p$ local orbitals were included. In the interstitial region, a plane-wave expansion with $R_{MT}K_{max}$ equal 8 was used for all the investigated systems, and the potential and the charge density were Fourier expanded with $G_{max}$=8.5. The modified tetrahedron method [14] was applied to integrate inside the Brillouin zone (BZ), and a k-point sampling with a 27×27×27 Monkhorst-Pack [15] mesh in the full BZ was considered as satisfactory for the cubic $Sc_3AlN$ and ScN systems, while for the hexagonal Sc a mesh of 30×30×32 was used.

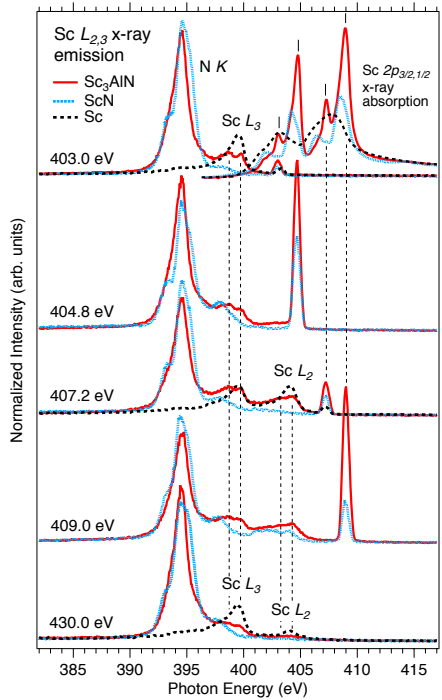

**Figure 3:** (Color online) top right: Soft x-ray absorption (SXA) spectra and left: resonant and nonresonant soft x-ray emission (SXE) spectra of the cubic inverse perovskite $Sc_3AlN$, ScN and Sc metal. The energy scale has been calibrated with the XPS binding energy of 404.9 eV for Sc $2p_{1/2}$ for ScN [2]. For comparison of the intensities, the spectra of the different systems were normalized to the calculated charge density occupations by the integrated areas while the SXA spectra were normalized by the step edge below and far above the Sc $2p$ absorption thresholds. All spectra were plotted on a photon energy scale (top) and a relative energy scale (bottom) with respect to the Fermi level ($E_F$). The SXE spectra were resonantly excited at the four Sc $2p$ absorption peak maxima at 403.0, 404.8, 407.2, and 409.0 eV and nonresonant far above the Sc $2p$ and N $1s$ absorption thresholds at 430.0 eV. The excitation energies of the SXE spectra are also indicated by the vertical ticks in the SXA spectra and by the strong elastic recombination peaks determining the total energy resolution (FWHM=0.65 eV).

Electronic band structure calculations were carried out using the experimental lattice parameters for *hcp*-Sc (a=3.31 Å; c=5.27 Å [16]) and *fcc*-ScN (a=4.50 Å [17]). For the cubic inverse perovskite $Sc_3AlN$, we optimized the unit cell volume with the APW+lo method and found a lattice constant of 4.37 Å, which is in reasonable agreement with the experimental value of a=4.40 Å[1]. Theoretical emission spectra were then evaluated at the converged ground-state density by multiplying the angular momentum projected density of states by the transition-matrix elements [1]. The electric-dipole approximation was employed so that only the transitions between the core states with orbital angular momentum $\ell$ to the $\ell \pm 1$ components of the electronic bands were considered. We previously successfully modelled SXE for other nitrides using the same type of theoretical scheme and obtained excellent agreement with experiment [18], giving validity and robustness to the present predictions for the $Sc_3AlN$ system. Finally, in order to interpret the experimental Sc $2p_{3/2,1/2}$ and N $1s$ SXA edges, the same kind of first-principles calculations were performed for the unoccupied states of cubic $Sc_3AlN$ and ScN structures. To account for the effect of the Sc $2p$ and N $1s$ core-holes on the unoccupied electronic band states, the excited atom was formally treated as an impurity inside a supercell of dimension 2×2×2 and a background charge was used to keep the systems neutral.

# 4 Results

## 4.1 Sc $L_{2,3}$ and N $K$ x-ray emission

Figure 3 shows Sc $L_{2,3}$ SXE spectra following the $3d4s \rightarrow 2p_{3/2,1/2}$ dipole transitions of Sc$_3$AlN (full curves), ScN (dotted curves), and Sc metal (dashed curves) excited at 403.0, 404.8, 407.2, 409.0, and 430.0 eV photon energies. As the occupied valence electronic structure of Sc metal nominally only contains one $3d$-electron, the $2p \rightarrow 3d$ SXA transition is strong and the SXE cross-section weak. SXA measurements (top, right curves) following the $2p_{3/2,1/2} \rightarrow 3d4s$ dipole transitions were used to determine the photon energies of the absorption peak maxima at the Sc $2p_{3/2}$ and $2p_{1/2}$ thresholds (vertical ticks). Note that the SXA spectrum of pure Sc metal (dashed curve) has single $2p_{3/2}$ and $2p_{1/2}$ absorption peaks. However, pure Sc metal is very sensitive to surface oxidation when radiated with x-rays and naturally form Sc$_2$O$_3$ and ScN at ambient pressure, which makes experiments relatively demanding in terms of vacuum quality and fast measurements, in particular for the surface sensitive TEY absorption. Surface oxidation from clean Sc metal was observed and monitored with TEY as a function of time after cleaning as a gradual build up of $t_{2g}$-$e_g$ ligand-field split double peaks. In the case of Sc$_3$AlN and ScN, the double peaks are observed as four narrow empty absorption bands in TEY, which are strongest in the case of Sc$_3$AlN, indicating a thin surface oxide layer and most empty $3d$ states. For Sc$_3$AlN and ScN, the two low-energy absorption peaks at 403.0 and 404.8 eV (1.75 eV $t_{2g}$-$e_g$ ligand-field splitting) are associated with the empty $2p_{3/2}$ core shell while the two high-energy peaks are associated with the empty $2p_{1/2}$ core shell.

The SXE spectra (to the left in Fig. 3) of Sc$_3$AlN and ScN are dominated by the N $2p \rightarrow 1s$ emission while the intensity of the $2p$ spin-orbit split Sc $3d \rightarrow 2p_{3/2,1/2}$ emission is weaker and appear at higher photon energy. The observed relative Sc/N intensity ratio is generally higher in the SXE spectra of Sc$_3$AlN than in ScN which is consistent with the fact that Sc$_3$AlN contains 10% more Sc than ScN. A low energy shoulder, more pronounced in ScN than in Sc$_3$AlN, is observed below the main N $K$ peak. The Rayleigh *elastic* scattering peaks of Sc$_3$AlN and ScN at the resonant excitation energies are strong which is a signature of localized electron excitations into narrow empty bands. Note that elastic scattering peaks are not only observed for Sc$_3$AlN and ScN, but weaker, also for pure Sc metal, which is generally not the case for non-correlated wide band transition metal systems in this experimental geometry [8]. The presence of elastic scattering in pure Sc metal is a signature of localized excitations and electron correlations as normally observed for rare earth materials [21] and transition metal oxides [22]. On the other hand, the absence of dispersing resonant *inelastic* x-ray scattering (RIXS) features following below the elastic peaks in all the SXE spectra is a signature of non-ionic systems with mainly delocalized electrons. Double-peak features with a splitting of 1.0 eV at constant emission energy (normal emission indicated by the vertical dashed lines in Fig. 3) are observed both at the Sc $L_3$ and $L_2$ emission of Sc$_3$AlN, but not in ScN and pure Sc metal. These spectral features can be associated with the hybridization with Al as discussed in section IV B.

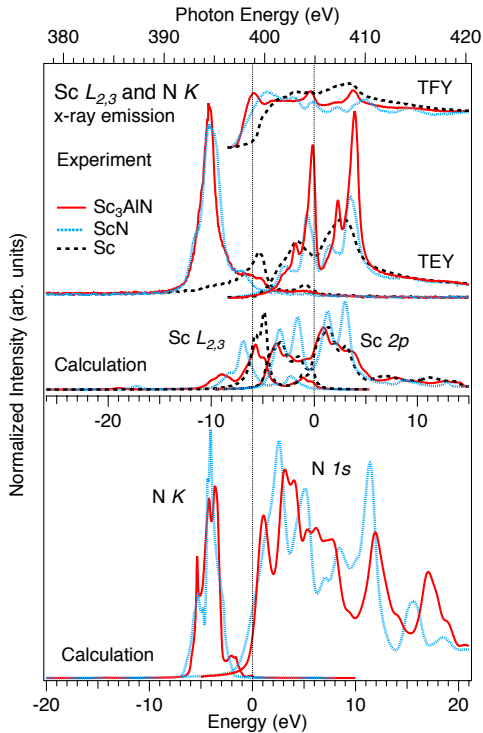

**Figure 4:** (Color online) Top: SXA (TFY and TEY) and nonresonant SXE spectra of Sc$_3$AlN, ScN and Sc metal excited at 430.0 eV (from Fig. 3). Middle: calculated Sc $L_{2,3}$ SXE and Sc $2p$ SXA spectra of Sc$_3$AlN, ScN, and Sc metal. Bottom: calculated N $K$ SXE and N $1s$ SXA spectra of Sc$_3$AlN and ScN. Note that the Sc $L_{2,3}$ and N $K$ emission partly overlap although the Sc $L_2$ and N $K$ Fermi level energies are separated by 6 eV. The integrated areas of the experimental and calculated spectra of the three systems were normalized to the sum of the calculated Sc $3d+4s$ and N $2p$ charge occupations of Sc$_3$AlN (Sc-$s$: 2.01e, Sc-$d$: 0.73e, N-$p$: 3.42e), fcc ScN (Sc-$s$: 2.01e, Sc-$d$: 0.66e, N-$p$: 3.41e) and hcp Sc metal (Sc-$s$: 2.03e, Sc-$d$: 0.78e) where the area of the Sc $L_2$ component was scaled down by the branching ratio and added to the $L_3$ component. In the calculated SXA spectra, core-hole correlation effects have also been included in the theory. To account for the difference in the calculated $1s$ core-level energies, the N $K$ SXE and N $1s$ SXA spectra of ScN were shifted by -1.8 eV, relative to the spectra of Sc$_3$AlN.

Although there are indications of electron correlations in all three Sc systems, the normal emission part of the SXE spectra appear as the electrons are mostly delocalized (wide bands) which makes electronic structure calculations suitable for the interpretation, particularly for nonresonant spectra. The observed nonresonant $L_3/L_2$ branching ratio of the Sc intensity is 3.0 for Sc$_3$AlN and is somewhat lower than for Sc metal (3.6), indicating rather good conduction properties for Sc$_3$AlN. Both for Sc$_3$AlN and Sc metal, the nonresonant $L_3/L_2$ branching ratio is higher than the statistical ratio (2:1) due to the additional Coster-Kronig process [8]. Moreover, the $L_3/L_2$ ratio is usually significantly higher than 2:1 for conducing systems than for insulators as the Coster-Kronig process is partly quenched by the band gap in insulators [23].

Figure 4 (top) shows SXA spectra (TFY and TEY) together with the non-resonant SXE spectra from the bottom of Fig. 3 (430.0 eV) in comparison to calculated Sc $L_{2,3}$ and N $K$ SXE as well as Sc $2p$ and N $1s$ SXA spectra (bottom) including dipole projected DOS and appropriate core-hole lifetime broadening of Sc$_3$AlN, ScN and Sc metal. Comparing the TEY and TFY measurements we note that the bulk sensitive fluorescence yield is more sensitive to the N absorption while the surface sensitive electron yield is more sensitive to Sc. The N $1s$ absorption peak maximum in TFY is identified at 398.9 eV which is 4.1 eV below the Sc $2p_{3/2}$ absorption maximum (403.0 eV) in TEY.

In the SXE spectra, the main Sc $3d$ - N $2p$ hybridization region in Sc$_3$AlN and ScN is identified in the energy region between -3 to -6 eV below E$_F$ with a peak at -4 eV. Due to the 30% lower N content, the experimental Sc$_3$AlN spectrum is narrower than for ScN. The N $K$ spectra of Sc$_3$AlN has a less pronounced shoulder at -5.5 eV than ScN due to hybridization with the Al $3s$ states, while the Al $3p$ states mainly hybridize between -1 and -3 eV below E$_F$. For comparison, the main bonding peak structure is observed at -4.8 eV below E$_F$ in Ti$_2$AlN [4]. The peak shift to higher energy from the E$_F$ in Sc$_3$AlN compared to Ti$_2$AlN indicates somewhat weaker interaction and bonding, although this is also governed by the relative intensities. The theoretically predicted weak feature due to Sc $3d$ - N $2s$ hybridization at -17 eV; however, does not appear experimentally as the N $2s$ states are effectively dipole forbidden. For pure Sc metal, a small shoulder observed 394 eV emission energy (-10 eV below $E_F$) is not predicted by one-electron theory and may be associated with many-body effects.

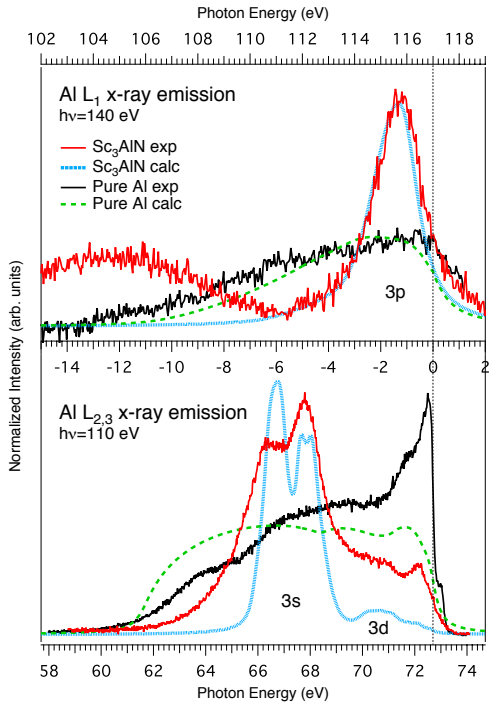

**Figure 5:** (Color online) Experimental (full curves) and calculated Al $L_1$ (dotted curves) and Al $L_{2,3}$ (dashed curves) SXE spectra of $Sc_3AlN$ and single-crystal Al(100). The experimental spectra were excited nonresonantly at 140 eV and 110 eV, respectively. The vertical dotted line indicates the Fermi level ($E_F$). A common energy scale with respect to the $E_F$ is indicated in the middle. For comparison of the peak intensities and energy positions, the integrated areas of the experimental and calculated spectra of the two systems were normalized to the calculated Al $3p$ and $3d+3s$ charge occupations of $Sc_3AlN$: ($3p$: 6.407e, $3d$: 0.015e, $3s$: 0.437e,) and pure Al metal ($3p$: 0.526e, $3d$: 0.090e, $3s$: 0.592e). The area for the $L_2$ component was scaled down by the experimental branching ratio and added to the $L_3$ component.

Note that for Sc metal, the $3d$ states dominate at the $E_F$, while for the semiconducting ScN, the Sc $3d$ states are at lower energy. From the combination of the spectral edges of the SXA and SXE spectra, the bandgap in ScN was estimated to 1.65 ± 0.5 eV. This is in fairly good agreement with reported indirect bandgaps of 0.9 ± 0.1 [24] and 1.3 ± 0.3 eV [25] in the literature. For $Sc_3AlN$ and Sc metal, the SXE spectra show that the bandgap is closed with Sc $3d$ states at the $E_F$ indicating good conduction properties, as demonstrated for $Sc_3AlN$ [1].

The observed Sc $2p$ spin-orbit splitting in SXE is 4.70 eV while it is somewhat smaller in SXA (4.25 eV) and in $2p$ core XPS (4.5 eV). Our calculated *ab initio* value of 4.64 eV is here in good agreement with SXE (4.70 eV), but is somewhat larger than for XPS. The calculated *ab initio* values of the spin-orbit splittings in band-structure calculations are usually underestimated for the early transition metals such as Ti and V and overestimated for the late transition metals such as Ni and Cu [8]. The reason for the difference in agreement between the experimental and theoretical spin-orbit splittings of different transition metals is not presently known, but must represent effects beyond effective one-electron theory, e.g., many-body effects.

## 4.2 Al $L_1$ and $L_{2,3}$ x-ray emission

Figure 5 shows Al $L_1$ (top panel) and Al $L_{2,3}$ (bottom panel) SXE spectra of $Sc_3AlN$ and Al metal [4], following the $3p \rightarrow 2s$ and $3s, 3d \rightarrow 2p_{3/2,1/2}$ dipole transitions, respectively. The measurements were made nonresonantly at 140 eV and 110 eV photon energies. Calculated spectra with the dipole projected pDOS and appropriate core-hole lifetime broadening are shown by the dotted and dashed curves.

The general agreement between experiment and theory is better for the $L_1$ emission involving spherically symmetric 2s core levels than for the $L_{2,3}$ emission involving 2p core levels. The $L_1$ fluorescence yield is much lower than the $L_{2,3}$ yield making the $L_1$ measurements more demanding in terms of beamtime. The main Al $L_1$ emission peak in Sc$_3$AlN between -1 and -2 eV on the common energy scale is due to Al 3p orbitals hybridizing with the Sc 3d orbitals. On the contrary, the weak $L_1$ emission of pure Al metal is very broad and flat without any narrow peak structures in the whole energy region, in agreement with our calculated $L_1$ spectrum. The very broad spectral structure experimentally observed below -6 eV in the $L_1$ spectrum of Sc$_3$AlN is, however, not reproduced in the calculated $L_1$ spectrum. It can be attributed to hybridization with N 2s and Sc 3d orbitals at the bottom of the valence band or additional background due to slightly enhanced reflectivity at the low-energy part of the spectrum which become visible at long data collection times. Compared to the spectra of Al metal, the spectral structure of Sc$_3$AlN is focussed to a specific energy region (-1.4 eV), as a consequence of Al hybridization with Sc and N. Bonding Al 3p orbitals appear at -1.4 eV below $E_F$ with Sc 3d orbitals (and -17 eV with N 2s orbitals). Comparing Sc$_3$AlN to Ti$_2$AlN [4], the shift of the N 2p orbitals to lower energy (from -4.0 eV to -4.8 eV) implies a shift of the 3d pDOS towards lower energy which also affects the spectral distributions of the Al $L_1$ and $L_{2,3}$ spectra [4].

The measured Al $L_{2,3}$ SXE spectrum of Sc$_3$AlN in the lower panel of Fig. 5 is dominated by $3s \rightarrow 2p_{3/2,1/2}$ dipole transitions while additional $3d \rightarrow 2p_{3/2,1/2}$ transitions mainly occur close to the $E_F$. The electronic structure of the Al atoms is strongly modified by the neighboring atoms. In particular, this is evident when comparing to the Al $L_{2,3}$ spectrum of Al metal where a very sharp peak has its maximum at -0.22 eV due to an edge resonance [26]. The small shoulder at +0.24 eV above $E_F$ is due to Al $L_2$ emission. We find the Al $L_3/L_2$ branching ratio of Al metal (4.15:1) to be larger than in the case of Sc metal (3.6:1). The 2p spin-orbit splitting is 0.46 eV, slightly larger than our calculated *ab initio* spin-orbit splitting of 0.44 eV. In contrast to the $L_{2,3}$ SXE spectrum of pure Al metal, the Al $L_{2,3}$ spectrum of Sc$_3$AlN has a strongly modified spectral weight towards lower emission energy. The main peak has its maximum at -5 eV below $E_F$ and a low-energy shoulder at -6 eV, indicating hybridization with the N 2p orbitals. We note that the energy positions of the -5 and -6 eV peaks in the calculated spectra are in good agreement with the experiment while the intensity distribution is reversed. The peaks at -0.5 and -1.5 eV are due to hybridization with the Sc 3d orbitals observed as double structures in Fig. 3 (vertical dashed lines). The Al 2p spin-orbit splitting is not resolved in Sc$_3$AlN. The partly populated 3d states are effectively withdrawn from the $E_F$ by charge transfer and form the peak structure at -0.5 eV from hybridization with Sc. For the calculated Al $L_{2,3}$ SXE spectra, the $3s,3d \rightarrow 2p_{3/2,1/2}$ matrix elements are important for the spectral functions as the intensity from the pDOS at the top of the valence band is enhanced and at the bottom reduced. However, the main disagreement between the experimental edge resonance and the one-electron calculation (especially at threshold) can be attributed to many-body effects, addressed in the Mahan-Nozières-De Dominicis (MND) theory of edge singularities [27, 28].

# 5 Discussion

Comparing the cubic inverse perovskite crystal structure of $Sc_3AlN$ in Fig. 1 with fcc ScN, it is clear that the physical properties and the electronic structure are affected by the introduction of the Al atoms. Intuitively, the conductivity should increase since Al metal is a good conductor with very high spectral intensity at the $E_F$ as shown in Fig. 5. The Sc $L_{2,3}$ SXE spectra in Fig. 3 show that the intensity at the $E_F$ is high in $Sc_3AlN$ and Sc metal due to the occupied Sc $3d$ pDOS at the $E_F$ indicating metallic-like properties while it is withdrawn in ScN below the bandgap. For free (delocalized) electrons, the conductivity is proportional to the number of states at the $E_F$ and is largely governed by the metal bonding. Although signs of electron correlations are found in the studied materials, it is useful to compare the trend in the conductivity with the density of states at the $E_F$; $Sc_3AlN$: 0.50, ScN: 0.00, Sc: 1.90, TiN: 0.43, $Ti_2AlN$: 0.41, TiC: 0.12, and $Ti_2AlC$: 0.34 states/eV/atom. This trend is in fairly good agreement with experimental values of the resistivity of $Sc_3AlN$ of 0.412 µΩm [1], which is slightly higher than for $Ti_2AlN$ films (0.39 µΩm [29]) and comparable to $Ti_2AlC$ (0.44 µΩm [30]). The resistivity of binary TiN is much lower (0.13 µΩm [31]) and for TiC more than an order of magnitude higher (2.50 µΩm [32]). ScN is a hard narrow band-gap semiconductor (very large resistivity) and pure Sc metal has a resistivity of 0.61 µΩm.

From Figs. 3-5, we distinguished three hybridization regions giving rise to the strong Sc $3d$ - N $2p$ bonding at -4 eV below $E_F$, weaker Sc $3d$ - Al $3p$ bonding at -1.4 eV and a predicted weak Sc $3d$ - N $2s$ interaction at -17 eV. The Sc $3d$ - N $2p$ covalent bonding and Sc $3d$ - N $2s$ hybridization are both deeper in energy from the $E_F$ than the Sc $3d$ - Al $3p$ hybridization indicating stronger bonding. Strengthening the relatively weak Sc $3d$ - Al $3p$ bonding would increase the stiffness of the material. The measured $E$-modulus of $Sc_3AlN$ is 249 GPa and for ScN it is 356 GPa which is somewhat lower than in the case of $Ti_2AlN$ (270 GPa [29]) and TiN (449 GPa [33]). By the specific choice of the elements in either a perovskite or an inverse perovskite crystal structure in the design of a material, the physical and mechanical properties can be tailored for specific applications. Further development of future stable cubic inverse perovskite materials include $Sc_3GaN$. Substitution of Al in $Sc_3AlN$ to Ga in $Sc_3GaN$ would nominally conserve the charge in the $sp$ valence band electronic structure. However, as in the case of Ge [34], the shallow $3d$ core level, at -19 eV below $E_F$ in Ga will effectively interact and withdraw charge density from the $sp$ valence band which is expected to influence the hybridization and chemical bonding in the material. This will be the subject of futher investigations.

# 6 Conclusions

In summary, we have investigated the electronic structure of the cubic inverse perovskite $Sc_3AlN$ in comparison to ScN and Sc metal. The combination of bulk sensitive and element selective soft x-ray emission spectroscopy and electronic structure calculations show that in $Sc_3AlN$ and ScN, the main Sc $3d$ - N $2p$ bond region appear -4 eV below $E_F$. In Al $L_1$ emission, a peak observed -1.4 eV below the $E_F$ is due to Al $3p$ states hybridized with Sc $3d$ states in a weaker covalent bonding. Our measured Al $L_{2,3}$ emission in $Sc_3AlN$ as compared to pure Al metal shows a significant shift of the main $3s$ intensity from the top of the valence band to hybridization regions at -5 and -6 eV. This signifies an effective transfer of charge from the Al $3d$ orbitals to the N $2p$ and Sc $3d$ orbitals. Absorption and emission measurements of clean Sc metal are presented and the measurements were found to be sensitive to surface oxidation while being irradiated by x-rays. The absorption and emission spectra of the investigated Sc systems reveal signs of electron correlations although the main part of the spectra appear as for delocalized and non-ionic systems. The revealed hybridization and chemical bond regions for ScAlN and ScN determine the electrical and thermal conductivity and the elastic properties of the materials.

# 7 Acknowledgements

We would like to thank the staff at MAX-lab for experimental support. This work was supported by the Swedish Research Council Linnaeus Grant LiLi-NFM, the Göran Gustafsson Foundation, the Swedish Strategic Research Foundation (SSF), Strategic Materials Reseach Center on Materials Science for Nanoscale Surface Engineering (MS$^2$E), and the Swedish Agency for Innvovation Systems (VINNOVA) Excellence Center on Functional Nanostructured Materials (FunMat). One of the authors (M.M.) wishes to acknowledge the Spanish Ministry of Science and Technology (MCyT) for financial support through the *Ramón y Cajal* program.